\documentclass[cits]{PoS}
\usepackage{amsmath, amsthm, amssymb}
\usepackage{array}

\newcommand{\eps}{\epsilon}
\newcommand{\taubar}{\bar\tau}
\def\rd{\mathrm{d}}

\title{Automated Calculation of Dijet Soft Functions\\ 
in Soft-Collinear Effective Theory\thanks{We are grateful to G.~Heinrich and T.~Hahn for helpful discussions. The work of GB and RR is supported by the Royal Society. JT acknowledge support from Hertford College, Oxford. Preprint: SI-HEP-2015-27, OUTP-15-30P.}}

\ShortTitle{Automated Calculation of Dijet Soft Functions in SCET}

\author{Guido Bell\\
        Rudolf Peierls Centre for Theoretical Physics, University of Oxford, 1 Keble Road, OX1 3NP, Oxford, United Kingdom\\ 
Theoretische Physik 1, Naturwissenschaftlich-Technische Fakult\"at,
Universit\"at Siegen, Walter-Flex-Strasse 3, 57068 Siegen, Germany\\
        E-mail: \email{bell@physik.uni-siegen.de}}

\author{Rudi Rahn\\
        Rudolf Peierls Centre for Theoretical Physics, University of Oxford, 1 Keble Road, OX1 3NP, Oxford, United Kingdom\\
        E-mail: \email{rudi.rahn@physics.ox.ac.uk}}

\author{\speaker{Jim Talbert}\\
        Rudolf Peierls Centre for Theoretical Physics, University of Oxford, 1 Keble Road, OX1 3NP, Oxford, United Kingdom\\
        E-mail: \email{jim.talbert@physics.ox.ac.uk}}       

\abstract{
We present a systematic framework for the calculation of dijet soft functions, which are defined as a matrix element of two light-like Wilson lines that is weighted by a generic measurement function. Based on a universal parametrisation of the phase-space integrals, we isolate the singular terms in Laplace space and evaluate the observable-dependent integrations numerically. We discuss the implementation of our algorithm into the public program {\tt SecDec}, and present results for various SCET-1 soft functions to two-loop order. In particular, we obtain new predictions for the C-parameter and the angularity variables. Finally, we discuss ongoing extensions and the future outlook of the project.
}

\FullConference{12th International Symposium on Radiative Corrections (Radcor 2015) and LoopFest XIV (Radiative Corrections for the LHC and Future Colliders)\\
                 15-19 June  2015\\
                 UCLA Department of Physics \& Astronomy Los Angeles, CA, USA}

\begin{document}

\section{Introduction}

Precision studies at collider experiments require a thorough understanding of strong-inter\-action effects. Whenever the QCD radiation is confined to jet-like configurations, the perturbative expansion in the strong coupling is spoiled by large logarithmic corrections. These Sudakov logarithms have to be resummed to all orders, which can be achieved e.g.~on the basis of effective field theory methods. The relevant effective theory is called Soft-Collinear Effective Theory (SCET) \cite{Bauer:2000yr,Bauer:2001yt,Beneke:2002ph}, which allows one to systematically disentangle the effects from hard, collinear and soft QCD radiation. 

In this work we consider dijet soft functions, which encode the effects from soft radiation in processes with two back-to-back jets. We aim at computing the soft functions to next-to-next-to-leading order (NNLO) in the perturbative expansion. Specifically, we will compute the two-loop soft anomalous dimension as well as the finite term of the renormalised soft function, which are important ingredients for next-to-next-to-leading logarithmic (NNLL) and N$^3$LL resummation, respectively.

In the last few years the computation of soft functions has attracted considerable attention. Whereas previous NNLO calculations \cite{Belitsky:1998tc,Kelley:2011ng,Monni:2011gb,Li:2011zp,Kelley:2011aa,Becher:2012za,Ferroglia:2012uy,Becher:2012qc,Czakon:2013hxa,vonManteuffel:2014mva,Boughezal:2015eha,Echevarria:2015byo} were performed analytically on a case-by-case basis, we for the first time present a systematic method for the automated (numerical) evaluation of generic dijet soft functions. Similar efforts for the automated extraction of hard functions are under way \cite{Becher:2014aya}, which may ultimately lead to fully automated resummations in SCET. Automated resummations for generic jet observables have so far only been performed with QCD techniques to NLL accuracy \cite{Banfi:2004yd}. The method has recently been extended to NNLL for $e^+e^-$ observables \cite{Banfi:2014sua}.

\section{Dijet soft functions}

We are concerned with dijet soft functions, which can generically be written in the form
\begin{equation}
S(\tau, \mu) = \frac{1}{N_c} \; \sum_X \; 
\mathcal{M}(\tau;\lbrace k_{i} \rbrace)\;
\mathrm{Tr}\; 
\langle 0 | S^{\dagger}_{\bar{n}} S_{n} | X \rangle
\langle X | S^{\dagger}_{n} S_{\bar{n}} | 0 \rangle \,,
\end{equation}
where $S_{n}$ and $S_{\bar{n}}$ are soft Wilson lines extending along two light-like directions $n^{\mu}$ and $\bar{n}^{\mu}$ with $n\cdot{\bar{n}}=2$. The definition involves a trace over colour indices and a generic measurement function $\mathcal{M}(\tau,\lbrace k_{i} \rbrace)$ that provides a constraint on the soft radiation $\lbrace k_{i} \rbrace$ according to the observable under consideration. In order to avoid complications with distribution-valued expressions, we assume that the measurement function is formulated in Laplace (or Fourier) space. We will mostly focus on $e^+e^-$ event-shape variables in the following, but the definition also includes (0-jet) hadron collider soft functions with an appropriate time (or anti-time) ordering of the Wilson lines.

In dimensional regularisation (DR) with $d=4-2\eps$, the matrix element contains explicit divergences from virtual corrections as well as implicit divergences from real radiation (which become explicit after integration over phase space). The important point to note is that the structure of the implicit divergences is independent of the observable. We may therefore isolate the divergences with a universal phase-space parametrisation, and compute the observable-dependent coefficients in the $\eps$-expansion numerically. Beyond one-loop order the phase-space integrals contain overlapping divergences, which we disentangle with a sector decomposition strategy \cite{Binoth:2000ps}.

\section{NLO calculation}

At tree level only the vacuum state contributes, and we assume that the measurement function is normalised to one. The soft function can up to two-loop order be written as
\begin{equation}
S(\tau, \mu) = 1 + \left(\frac{Z_\alpha\alpha_s}{4\pi}\right) (\mu^2 \taubar^2)^\eps \; S_1(\eps) + \left(\frac{Z_\alpha\alpha_s}{4\pi}\right)^2 (\mu^2 \taubar^2)^{2\eps} \; S_2(\eps) + \mathcal{O}(\alpha_s^3)\,,
\end{equation}
where $\taubar = \tau e^{\gamma_E}$ and $\alpha_s$ is the $\overline{\textrm{MS}}$-renormalised coupling, which is related to the bare coupling constant $\alpha_s^0$ via $Z_\alpha \alpha_s\,\mu^{2\eps}=e^{-\eps\gamma_E}(4\pi)^\eps \alpha_s^0$ with $Z_\alpha = 1-\beta_0\alpha_s/(4\pi\eps)$ and $\beta_0 = 11/3\, C_A - 4/3\,T_F n_f$. At NLO the computation involves one-loop virtual corrections to the vacuum state and one-gluon real emission diagrams. The virtual corrections are scaleless and vanish in DR. Denoting the gluon momentum by $k^\mu$, the real emission contribution takes the form
\begin{equation}
S_1(\eps)  = \frac{(4\pi e^{\gamma_E} \tau^2)^{-\eps}}{(2\pi)^{d-1}} \,
\int d^{d}k \;\, \delta(k^{2}) \,\theta(k^{0}) 
\,\mathcal{M}(\tau; k) \, |\mathcal{A}(k)|^{2} \,.
\end{equation}
At NLO the squared matrix element is given by
\begin{equation}
| \mathcal{A}(k) |^{2} = \frac{64 \pi^2 C_{F}}{k_{+}k_{-}}
\end{equation}
with $k_+ = n \cdot k$ and $k_- = \bar n \cdot k$. In order to disentangle the singularity structure, we split the integration region into two hemispheres with $k_->k_+$ (left) and $k_+>k_-$ (right). In the left hemisphere we substitute
\begin{equation}
k_{-} = \frac{k_{T}}{\sqrt{y}}\,, \qquad\qquad k_{+} = k_{T} \sqrt{y}\,,
\end{equation}
in terms of the magnitude of the transverse momentum $k_{T}=\sqrt{k_+ k_-}$ and a measure of the rapidity $y=k_+/k_-$. We then impose the principal assumption of our approach, namely that the NLO measurement function can be written in the generic form
\begin{equation}
\label{eq:measure:NLO}
\mathcal{M}(\tau; k) = \exp\big(-\tau\, k_{T}\, y^{n/2}\, f(y,\theta)\big)\,,
\qquad\quad y\in\{0,1\}\,.
\end{equation}
The exponential arises from the Laplace transformation of the momentum-space measurement function. We assume that the Laplace variable $\tau$ has the dimension $1/$mass, which fixes the linear dependence on $k_T$ on dimensional grounds. The measurement function may further have a non-trivial angular dependence, since the measurement may not necessarily be performed with respect to the jet axis. If so, we project the measurement vector $v^\mu$ onto the transverse plane and introduce $\theta$ as the angle between $v_\perp^\mu$ and $k_\perp^\mu$. Finally, the power $n$ is fixed by the requirement that the function $f(y,\theta)$ is finite and non-zero in the collinear limit $y\to0$. Table \ref{tab:measure:NLO} reveals that the form (\ref{eq:measure:NLO}) is a general ansatz, as many observables of varying degrees of complexity fall within its domain.
\begin{table}
\center
\setlength{\extrarowheight}{10pt}
\scalebox{.84}{\begin{tabular}{|c|c|c|}
\hline \hline
Soft function & $n$ & $f(y,\theta)$ \\[6pt]
\hline \hline
Thrust \cite{Fleming:2007qr} & $1$ & $1$ \\[6pt]
\hline
Angularities \cite{Hornig:2009vb} & $1-A$ & $1$ \\[6pt]
\hline
Recoil-free broadening \cite{Larkoski:2014uqa} & $0$ & $1/2$  \\[6pt]
\hline
C-parameter \cite{Hoang:2014wka} & $1$ & $1/(1+y)$ \\[6pt]
\hline
Threshold Drell-Yan \cite{Korchemsky:1993uz} & $-1$ & $1+y$  \\[6pt]
\hline
W@large $p_{T}$ \cite{Becher:2009th} & -1 & $1 + y - 2 \sqrt{y}\, \cos\theta$ \\[6pt]
\hline
$e^+e^-$ transverse thrust \cite{Becher:2015gsa} & 1 & $\frac{1}{s\sqrt{y}} 
\bigg( \sqrt{ (c \cos\theta +\Big( \frac{1}{\sqrt{y}} - \sqrt{y} \Big) \frac{s}{2} \Big)^2 + 1 - \cos^2\theta} - \Big|c \cos\theta +\Big( \frac{1}{\sqrt{y}} - \sqrt{y} \Big) \frac{s}{2}\Big| \bigg) $ \\[6pt]
\hline
\end{tabular}}
\caption{Sample soft functions that fall within our ansatz of the measurement function (3.5), and their corresponding expressions for $n$ and $f(y,\theta)$. In the last line we used $s=\sin\theta_B$ and $c=\cos\theta_B$, where $\theta_B$ is the angle between the beam and the jet axis.}
\label{tab:measure:NLO}
\end{table} 

In the right hemisphere we proceed similarly with the role of $k_-$ and $k_+$ interchanged. Assuming that the measurement function is symmetric under the exchange of $n\leftrightarrow\bar n$, we arrive at the following master formula for the calculation of NLO dijet soft functions
\begin{equation}
\label{eq:NLO:master}
S_1(\eps) = \frac{8C_F\,e^{-\gamma_E\eps}}{\sqrt{\pi}}\,\frac{\Gamma(-2\eps)}{\Gamma(1/2-\eps)}\;\int_0^1 dy \;\, y^{-1+n\eps}\;
\int_{-1}^1 d\cos\theta \;\,\sin^{-1-2\eps}\theta \;\;
[f(y,\theta)]^{2\eps}\,.
\end{equation}
As desired, the singularity structure of the soft function has been completely factorized. The soft singularity in the limit $k_T\to0$ gives rise to the factor $\Gamma(-2\eps)$, and the collinear singularity in the limit $y\to0$ is encoded in the factor $y^{-1+n\eps}$ (note that the function $f(y,\theta)$ is finite by construction in this limit). We further observe that the collinear singularity is not regularised for $n=0$, which corresponds to a SCET-2 soft function. It is well understood that SCET-2 observables require an additional rapidity regulator, which can easily be implemented in our approach in the form proposed e.g.~in \cite{Becher:2011dz}. 

\section{NNLO calculation}

At NNLO the computation involves two-loop virtual, mixed real-virtual and double real emission diagrams. The two-loop virtual corrections are again scaleless and vanish in DR. The matrix element of the real-virtual corrections reads
\begin{equation}
| \mathcal{A}_{RV}(k) |^{2} = - 64 \pi^2 \; C_A C_F \;\,
\frac{\pi^2\,\Gamma(-\eps)\,\cot(\pi\eps)}{\Gamma(-2\eps)\,\sin(\pi\eps)}
\;\;k_{+}^{-1-\eps}k_{-}^{-1-\eps}\,.
\end{equation}
As the structure is similar to the NLO calculation, we can procced along the same lines and obtain
\begin{align}
S_{2}^{RV}(\eps)  
&= -8C_AC_F\,e^{-2\gamma_E\eps} \;\,
\frac{\pi^{3/2}\, \Gamma(-\eps)\, \Gamma(-4 \eps)\, \cot(\pi\eps)}{\Gamma(-2 \eps)\,\Gamma(1/2 - \eps)\,\sin(\pi\eps)} 
\nonumber\\[0.3em]
&\qquad\times\;
\int_0^1 dy \;\, y^{-1+2n\eps}\;
\int_{-1}^1 d\cos\theta \;\,\sin^{-1-2\eps}\theta \;\;
[f(y,\theta)]^{4\eps}\,.
\end{align}
The matrix element of the double real emission contribution consists of three colour structures: $C_{F}^{2}$, $C_{F}C_{A}$ and $C_{F}T_{F}n_{f}$. We assume that the measurement function is consistent with non-abelian exponentiation \cite{Gatheral:1983cz,Frenkel:1984pz}, 
\begin{equation}
\label{eq:measure:NAE}
\mathcal{M}(\tau; k,l) =\mathcal{M}(\tau; k) \; \mathcal{M}(\tau;l) \,,
\end{equation}
and so the $C_{F}^{2}$ contribution can be expressed in terms of the NLO expression (\ref{eq:NLO:master}). For the remaining colour structures, we start from
\begin{equation}
S_2^{RR}(\eps)  = \frac{(4\pi e^{\gamma_E} \tau^2)^{-2\eps}}{(2\pi)^{2d-2}} \,
\int d^{d}k \;\, \delta(k^{2}) \,\theta(k^{0}) \,
\int d^{d}l \;\, \delta(l^{2}) \,\theta(l^{0}) 
\;\mathcal{M}(\tau; k,l) \, |\mathcal{A}_{RR}(k,l)|^{2} \,.
\end{equation}
The squared matrix element of the $C_{F}T_{F}n_{f}$ contribution is given by
\begin{equation}
|\mathcal{A}_{RR}(k,l)|^{2} = 2048 \pi^{4} \,C_{F} T_{F} n_{f} \;\, 
\frac{2k\cdot{l} \, (k_{-} + l_{-}) \, (k_{+} + l_{+}) - (k_{-}l_{+}-k_{+}l_{-})^{2}}{(k_{-} + l_{-})^{2}\,(k_{+} + l_{+})^{2}\,(2k\cdot{l})^{2}} \,,
\end{equation}
and the corresponding expression for the $C_F C_A$ colour structure can be found e.g.~in \cite{Becher:2012qc}. Unlike the NLO case, the singularity structure is non-trivial and there exist overlapping divergences e.g.~in the limit $k_-\to0$ and $l_-\to0$.  In order to disentangle the singularity structure of the double real emission contribution, we parametrise the phase-space integrals in terms of the variables
\begin{align}
p_{-} &= k_{-} + l_{-} \,, \qquad\qquad 
a = \sqrt{\frac{k_{-}l_{+}}{k_{+}l_{-}}} \,,
\nonumber\\[0.3em]
p_{+} &= k_{+} + l_{+} \,, \qquad\qquad
b =\sqrt{\frac{k_{-}k_{+}}{l_{-}l_{+}}} \,.
\end{align}
Here $p_-$ and $p_+$ are the total light-cone momenta, $a$ is a measure of the rapidity difference of the two emitted partons, and $b$ is the ratio of their transverse momenta. The matrix element in addition depends on the angle $\theta_{kl}$ between $k_\perp^\mu$ and $l_\perp^\mu$, and the measurement function may introduce two further angles between the measurement vector $v_\perp^\mu$ and $k_\perp^\mu$ ($\theta_k$), and $v_\perp^\mu$ and $l_\perp^\mu$ ($\theta_l$). As in the NLO calculation, we substitute $p_{T}=\sqrt{p_+ p_-}$ and $y=p_+/p_-$. We further assume that the NNLO measurement function can be cast into the form
\begin{equation}
\label{eq:measure:NNLO}
\mathcal{M}(\tau; k,l) = \exp\big(-\tau\, p_{T}\, y^{n/2}\, F(a,b,y,\theta_k,\theta_l)\big)\,,
\qquad\quad a,b,y\in\{0,1\}\,.
\end{equation}
The linear dependence on $p_T$ is again fixed on dimensional grounds, and the factor $y^{n/2}$ is a consequence of the property (\ref{eq:measure:NAE}) and the structure of the NLO measurement function (\ref{eq:measure:NLO}). The function $F(a,b,y,\theta_k,\theta_l)$ encodes the non-trivial dependence on the observable at NNLO, and it is finite and non-zero in the limit $y\to0$. One can easily derive the corresponding expressions for the sample soft functions in Table \ref{tab:measure:NLO}, but here we only quote the measurement function for $W$-production at large transverse momentum as an example,
\begin{equation}
F(a,b,y,\theta_k,\theta_l) = 
1 + y - 2 \;\sqrt{\frac{a y}{(1+a b)(a+b)}} \;
\big(b \cos\theta_k + \cos\theta_l\big)\,.
\end{equation}
In particular, we observe that 
\begin{equation}
F(a,b,y,\theta_k,\theta_l) \to f(y,\theta)
\end{equation}
in the soft limit $k^\mu\to0$, which corresponds to $b\to 0$ (and $\theta_l\to\theta$) in our parametrisation. The same is true in the limit in which the two partons become collinear with $a\to 1$ and $\theta_k\to\theta_l=\theta$. These scaling rules in the soft and collinear limits are not accidental, but are a reflection of infrared-collinear safety, and so they are universal properties of the NNLO measurement function. 

As the $p_T$-dependence is universal for the considered class of observables, we can perform this integration explicitly. We further use the symmetries in $n\leftrightarrow\bar n$ and $k\leftrightarrow l$ to map the integration region in $\{a,b,y\}$ onto the unit hypercube. This results in two contributions that involve the measurement functions
\begin{equation}
F(a,b,y,\theta_k,\theta_l)\,,\qquad\qquad
F(1/a,b,y,\theta_k,\theta_l)\,.
\end{equation}
Without going into further details here, we finally map the angular integrations onto the unit hypercube with a suitable transformation, $(\theta_k, \theta_l, \theta_{kl}) \to (t_k, t_{kl}, x_l)$. We are thus left with a six-dimensional integral representation of the double real emission contribution. Similar to the NLO formula (\ref{eq:NLO:master}), it contains an explicit singularity from $p_T\to0$ and an implicit divergence for $y\to0$. In addition, we find an overlapping divergence in the limit $a\to 1$ and $t_{kl}\to 0$ (which corresponds to $\theta_{kl}\to0$). The $C_{F}T_{F}n_{f}$ contribution thus starts with a $1/\eps^3$ pole. The same strategy can be applied for the $C_F C_A$ colour structure, which turns out to contain an extra divergence in the limit $b\to0$, and therefore starts with a $1/\eps^4$ contribution.

The integral representations that we have derived are amenable to the public program {\tt SecDec} \cite{Carter:2010hi,Borowka:2012yc,Borowka:2015mxa}. Its \emph{general} mode allows us to define the generic factors that contain all the implicit divergences in the main template file on which the sector decomposition algorithm operates. The observable-dependent measurement function, on the other hand, is kept symbolic during the sector decomposi\-tion and subtraction steps, and its explicit form is resolved only at the final numerical integration stage.

For the numerical integrations {\tt SecDec} offers interfaces to the Cuba library \cite{Hahn:2004fe} and Bases \cite{Kawabata:1995th}. We typically use Divonne and Cuhre as our default Cuba integrators, and use Bases for independent cross-checks. Both Bases and the Cuba library return error estimates, which we do not quote in the following, since we need to investigate further if they are trustworthy. For angular-independent observables, the integrations run over four variables and {\tt SecDec} produces results at six digit precision in a few hours on a single machine. For angular-dependent observables, on the other hand, the speed of convergence is significantly reduced and we typically obtain four digits in a day (still on a single machine). Further improvements on the numerical side are desirable and progress towards that goal will be reported in a future publication \cite{BRT}.

\section{Renormalisation}

The calculation we have outlined so far yields the bare soft function $S_0$. In 
Laplace space the soft function renormalises multiplicatively, $S = Z_S S_0$, and the renormalised soft function fulfils the renormalisation group (RG) equation
\begin{align}
\label{eq:RG}
\frac{\rd}{\rd \ln\mu}  \; S(\tau,\mu)
&= - \frac{1}{n} \,\bigg[ 4 \,\Gamma_{\mathrm{cusp}}(\alpha_s) \, 
\ln(\mu\taubar) 
-2 \gamma^{S}(\alpha_s) \bigg] \; S(\tau,\mu) \,.
\end{align}
Here $\Gamma_{\mathrm{cusp}}(\alpha_s)$ denotes the cusp anomalous dimension and $\gamma^{S}(\alpha_s)$ is the soft anomalous dimension. We find it convenient to define the anomalous dimensions with a common prefactor $(-1/n)$, where $n$ reflects the scaling of the observable in the soft-collinear limit according to (\ref{eq:measure:NLO}). 
Expanding $\Gamma_{\mathrm{cusp}}(\alpha_s)= \sum_{n=0}^\infty \,\Gamma_n (\frac{\alpha_s}{4\pi})^{n+1}$ and $\gamma^{S}(\alpha_s) = \sum_{n=0}^\infty \,\gamma^{S}_n \,(\frac{\alpha_s}{4\pi})^{n+1}$, the two-loop solution of the RG equation takes the form
\begin{align}
S(\tau,\mu) &= 
1 + \left( \frac{\alpha_s}{4 \pi} \right) 
\left\{ -\frac{2\Gamma_0}{n} \,L^2 
+ \frac{2\gamma_0^S}{n} \,L 
+ c_1^S \right\}
+\left( \frac{\alpha_s}{4 \pi} \right)^2 
\bigg\{ \frac{2 \Gamma_0^2}{n^2} L^4 
-  4\Gamma_0\left( \frac{\gamma_0^S}{n^2} + \frac{\beta_0}{3n} \right) 
L^3 
\nonumber\\[0.3em]  
 &\quad
 - 2 \left( \frac{\Gamma_1}{n} - \frac{(\gamma_0^S)^2}{n^2} 
 - \frac{\beta_0 \gamma_0^S}{n} +\frac{\Gamma_0 c_1^S}{n} \right) L^2 
+ 2 \left (\frac{\gamma_1^S}{n} +\frac{\gamma_0^S c_1^S}{n} +\beta_0 c_1^S \right) L + c_2^S \bigg\} 
\end{align}
with $L=\ln(\mu\taubar)$. The $Z$-factor $Z_{S}$ fulfils the same RG equation (\ref{eq:RG}), and its explicit form is given to two-loop order by
\begin{align} 
Z_{S} &= 1 + \left( \frac{\alpha_s}{4 \pi} \right) 
\left\{ \frac{\Gamma_0}{n}\,\frac{1}{\eps^2} + 
\frac{2\Gamma_0 L - \gamma_0^S}{n}\, \frac{1}{\eps}
\right\}
+\left( \frac{\alpha_s}{4 \pi} \right)^2 
\bigg\{ \frac{\Gamma_0^2}{2n^2}\,\frac{1}{\eps^4}
+\Gamma_0 \left( \frac{2\Gamma_0}{n^2}\,L
-\frac{\gamma_0^S}{n^2}
 - \frac{3\beta_0}{4n}\right)\,\frac{1}{\eps^3}
\nonumber\\[0.3em]  
 &\quad
+\bigg( \frac{2\Gamma_0^2}{n^2}\,L^2 
- \Gamma_0 \Big(
\frac{2\gamma_0^S}{n^2} + \frac{\beta_0}{n} \Big)L
+\frac{\Gamma_1}{4n} + \frac{(\gamma_0^S)^2}{2n^2}
+ \frac{\beta_0\gamma_0^S}{2n}
\bigg) \, \frac{1}{\eps^2}
 + \frac{2\Gamma_1 L - \gamma_1^S}{2n}\, \frac{1}{\eps}
\bigg\}\,.
\end{align}
The leading expansion coefficients of the cusp anomalous dimension are
\begin{equation} 
\Gamma_0 = 4 C_F\,,
\qquad\quad
\Gamma_1 = 4 C_F
\left\{ \left(\frac{67}{9}-\frac{\pi^2}{3}\right) C_A
- \frac{20}{9} T_F n_f \right\}\,.
\end{equation}
The cancellation of the divergences $1/\eps^j$ with $j=2,3,4$ in the renormalised result provides a strong check of our calculation. We can then extract the anomalous dimensions $\gamma_0^S$ and $\gamma_1^S$ from the $1/\eps$ pole terms, and the coefficients $c_1^S$ and $c_2^S$ of the renormalised soft function from the finite terms of the two-loop calculation.

\section{Results}

We present results for all SCET-1 observables of Table \ref{tab:measure:NLO}, except for $e^+e^-$ transverse thrust. For these observables the NLO calculation as well as the NNLO mixed real-virtual correction are trivial and can be performed analytically. As the respective measurement functions are consistent with non-abelian exponentiation, the $C_F^2$ contribution is also known analytically. 

We thus use {\tt SecDec} to compute the $C_F C_A$ and $C_F T_F n_f$ double real emission contributions. We write the two-loop anomalous dimension and the finite term in the form
\begin{align} 
\gamma_1^S &= \gamma_1^{C_A} \;C_F C_A 
+ \gamma_1^{n_f} \;C_F T_F n_f\,,
\nonumber\\[0.3em]  
c_2^S &= c_2^{C_A} \;C_F C_A 
+ c_2^{n_f} \;C_F T_F n_f
+ \frac12 (c_1^S)^2\,.
\end{align}
Table \ref{tab:results} summarises our results for thrust, C-parameter, threshold Drell-Yan production and $W$-production at large transverse momentum. Strictly speaking, the soft function for $W$-production at large $p_T$ is not of the dijet-type considered here, but as argued in \cite{Becher:2012za} the diagrams with attachments to the third Wilson line are all scaleless and vanish up to NNLO. We can therefore consider this function as an example with a non-trivial angular dependence.\footnote{In this case, the colour structure is also slightly different with $C_F \to C_F -C_A/2$ in the $q\bar q\to g$ and $C_F \to C_A/2$ in the $q g\to q$ and $gg\to g$ channels, see \cite{Becher:2012za}.}
\begin{table}
\center
\setlength{\extrarowheight}{10pt}
\scalebox{.92}{\begin{tabular}{|c|c|c|c|c|c|c|}
\hline \hline
Soft function & $\gamma_0^S/C_F$ & $c_1^S/C_F$ 
& $\gamma_1^{C_A}$ & $\gamma_1^{n_f}$ 
& $c_2^{C_A}$ & $c_2^{n_f}$ \\[6pt]
\hline \hline
Thrust \cite{Kelley:2011ng,Monni:2011gb} & $0$ & $-\pi^2$ & 
$\begin{array}{c}
15.7945\\[-0.7em]
(15.7945)\\[0.5em]
\end{array}$
 & 
 $\begin{array}{c}
3.90981\\[-0.7em]
(3.90981)\\[0.5em]
\end{array}$
 & 
$\begin{array}{c}
-56.4992\\[-0.7em]
(-56.4990)\\[0.5em]
\end{array}$ 
& 
$\begin{array}{c}
43.3902\\[-0.7em]
(43.3905)\\[0.5em]
\end{array}$
\\[6pt]
\hline
C-parameter \cite{Hoang:2014wka} & $0$ & $-\pi^2/3$ & 
$\begin{array}{c}
15.7947\\[-0.7em]
(15.7945)\\[0.5em]
\end{array}$
 & 
 $\begin{array}{c}
3.90980\\[-0.7em]
(3.90981)\\[0.5em]
\end{array}$
 & 
$\begin{array}{c}
-57.9754\\[-0.7em]
(-)\\[0.5em]
\end{array}$ 
& 
$\begin{array}{c}
43.8179\\[-0.7em]
(-)\\[0.5em]
\end{array}$
\\[6pt]
\hline
Threshold Drell-Yan \cite{Belitsky:1998tc} & $0$ & $\pi^2/3$ & 
$\begin{array}{c}
15.7946\\[-0.7em]
(15.7945)\\[0.5em]
\end{array}$
 & 
 $\begin{array}{c}
3.90982\\[-0.7em]
(3.90981)\\[0.5em]
\end{array}$
 & 
$\begin{array}{c}
6.81281\\[-0.7em]
(6.81287)\\[0.5em]
\end{array}$ 
& 
$\begin{array}{c}
-10.6857\\[-0.7em]
(-10.6857)\\[0.5em]
\end{array}$
\\[6pt]
\hline
W@large $p_{T}$ \cite{Becher:2012za} & $0$ & $\pi^2$ & 
$\begin{array}{c}
15.88\\[-0.7em]
(15.7945)\\[0.5em]
\end{array}$
 & 
 $\begin{array}{c}
3.905\\[-0.7em]
(3.90981)\\[0.5em]
\end{array}$
 & 
$\begin{array}{c}
-2.78\\[-0.7em]
(-2.65010)\\[0.5em]
\end{array}$ 
& 
$\begin{array}{c}
-25.28\\[-0.7em]
(-25.3073)\\[0.5em]
\end{array}$
\\[6pt]
\hline
\end{tabular}}
\caption{Anomalous dimensions and finite terms of the renormalised soft function for sample SCET-1 observables. The upper numbers are the numerical results that we obtain with the {\tt SecDec} implementation of our algorithm, and the lower ones correspond to the known analytic expressions.}
\label{tab:results}
\end{table} 

The first three entries in Table \ref{tab:results} correspond to angular-independent measurement functions. For these observables we are left with four-dimensional numerical integrations that can be evaluated very accurately as can be seen from the comparison with the known analytic results. The finite term of the renormalised C-parameter soft functions has not been calculated so far, but a numerical extraction from a comparison with the {\tt EVENT2} generator has been performed in \cite{Hoang:2014wka}. The authors find
\begin{equation}
c_2^{C_A}= -58.16 \pm 0.26\,,\qquad\quad
c_2^{n_f}= 43.74 \pm 0.06\,\qquad\quad 
\text{(C-parameter)}
\end{equation}
which agrees well with our findings. Based on the numbers that we find for the thrust and threshold Drell-Yan soft functions -- which are of the same numerical complexity -- we are, however, led to expect that our numbers are significantly more accurate.

The measurement function for $W$-production at large transverse momentum has a non-trivial angular dependence. In this case we therefore face six-dimensional integrations, and from Table \ref{tab:results} we see that we typically loose two digits of precision for those observables. The situation is particularly bad for the coefficient $c_2^{C_A}$, and the problem can be traced back to a large cancellation between the double real emission and the mixed real-virtual contributions. For the finite term of the double real emission term, we find $108.62$, which is to be compared with the analytic result $108.75$. The accuracy of our number for the double real emission contribution is thus similar to the ones in the last line of Table \ref{tab:results}, but due to the large cancellation our final result for $c_2^{C_A}$ is poor\footnote{In the meantime we significantly improved our numerical routines, and we now (Dec 2015) obtain $c_2^{C_A}=-2.65034$ and $c_2^{n_f}=-25.3073$ in a few hours  on a single 8-core machine.}.

\begin{figure}[t]
\centering
\hspace{2mm}
\includegraphics[scale=0.55]{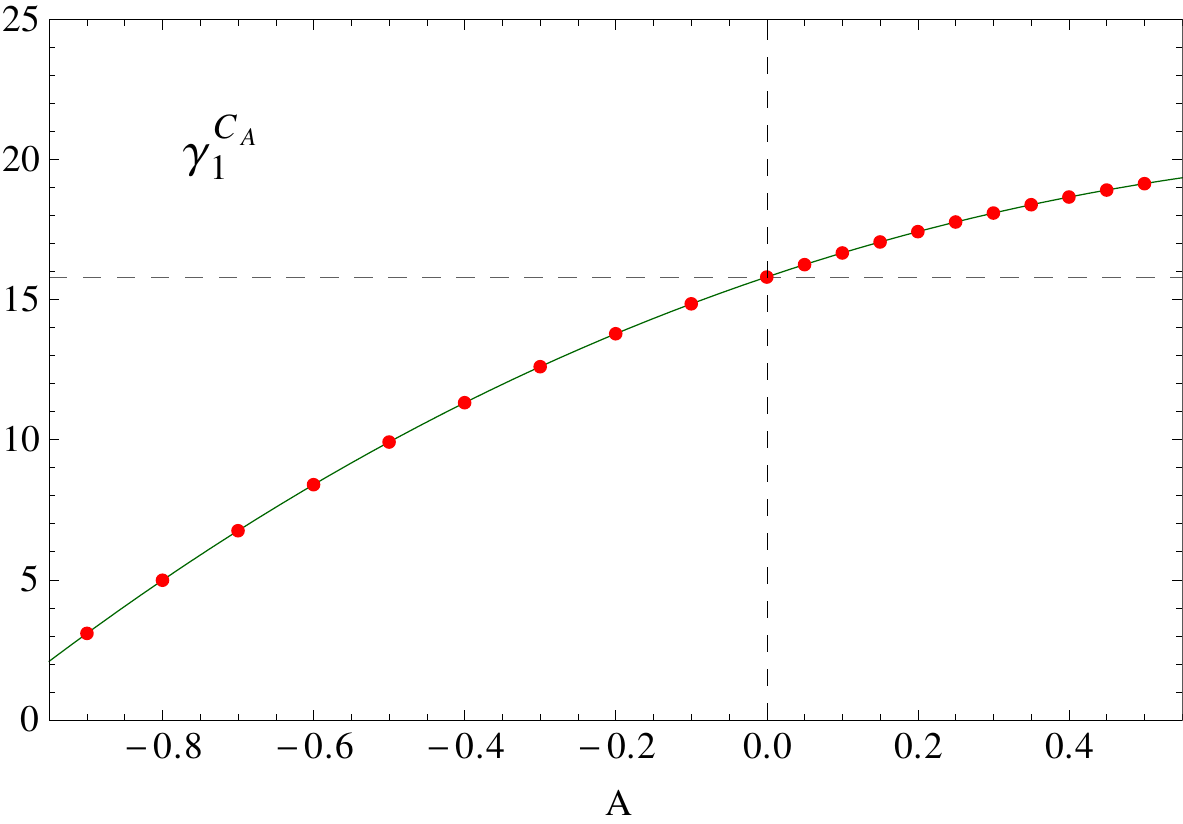}
\hspace{8mm}
\includegraphics[scale=0.55]{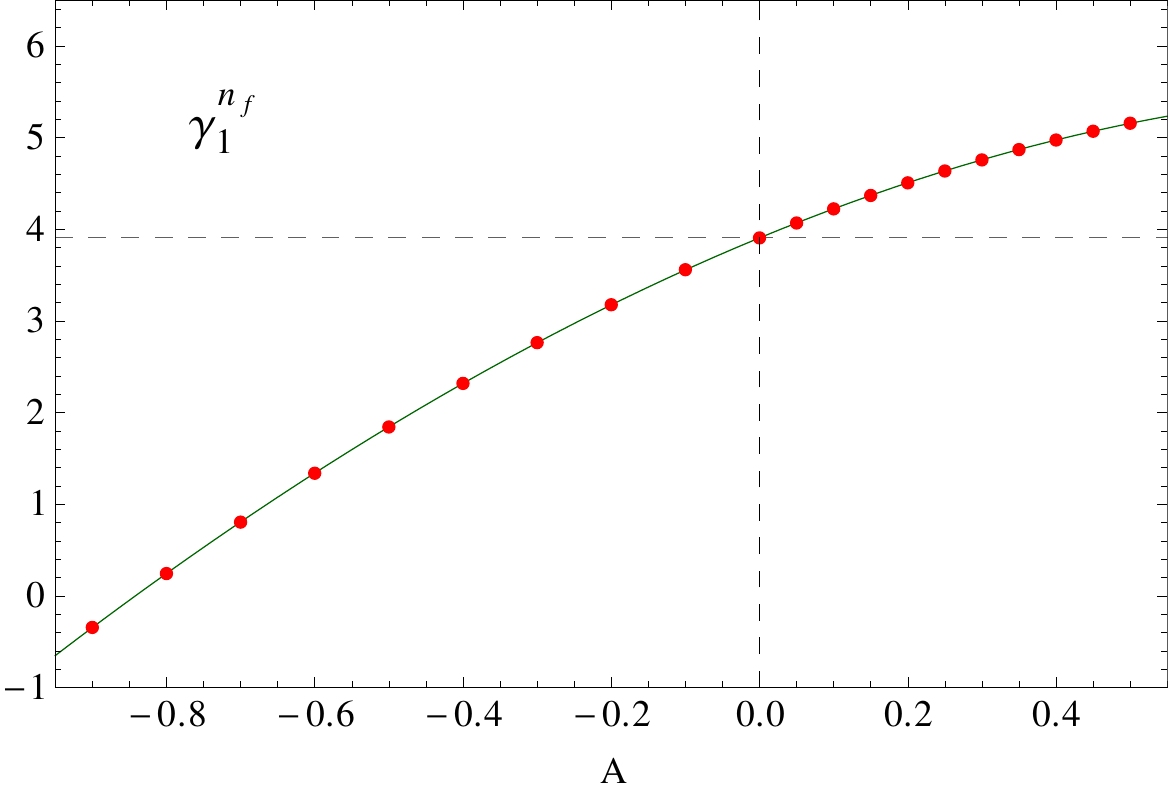}
\\[1em]
\includegraphics[scale=0.545]{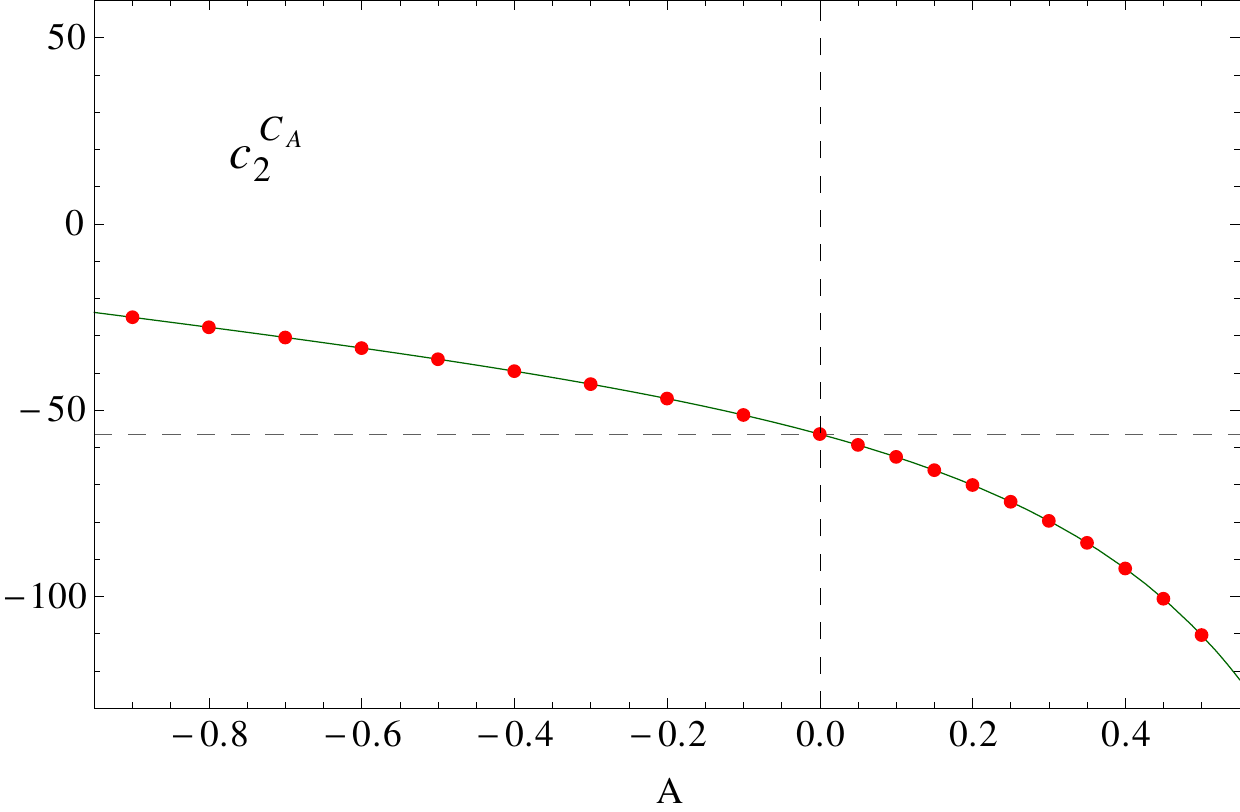}
\hspace{8mm}
\includegraphics[scale=0.52]{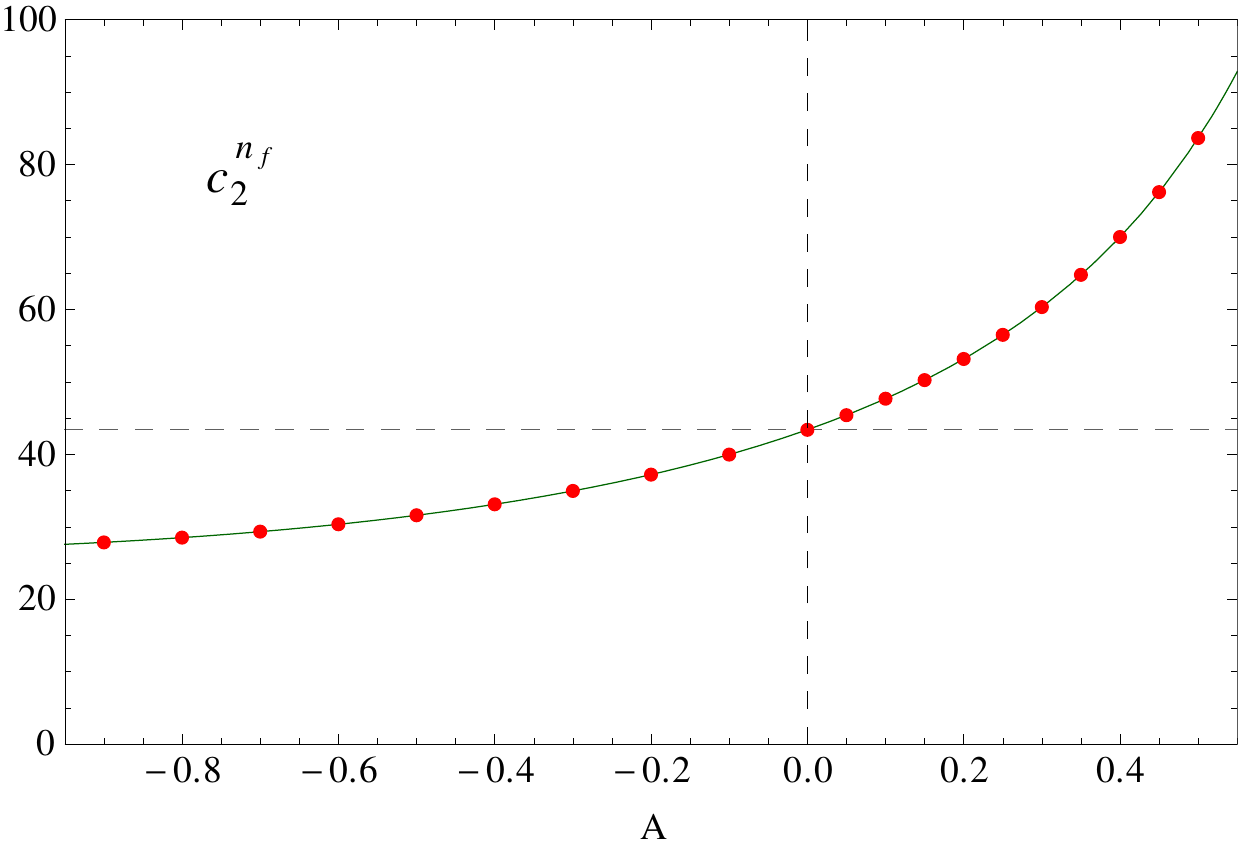}
\caption{Two-loop anomalous dimension and finite term of the renormalised angularity soft function. The dashed line indicates the analytic thrust number, and the green (solid) line represents a fit to the data points.}
\label{fig:angularities}
\end{figure}

We next consider the angularity soft function, which has only been computed to one-loop order so far. In this case the anomalous dimension and the finite term are functions of the angularity $A$, which interpolates between thrust (for $A=0$) and total broadening (for $A=1$). The one-loop ingredients are \cite{Hornig:2009vb}
\begin{equation}
\gamma_0^{S}(A) = 0\,,\qquad\quad
c_1^{S}(A) = - \frac{\pi^2}{1-A}\; C_F\,.\qquad\quad 
\text{(Angularities)}
\end{equation}
We have evaluated the two-loop soft function with our numerical techniques for twenty values of the angularity between $A=-0.9$ and $A=0.5$. The results for the two-loop anomalous dimension and the finite term of the renormalised soft function are displayed in Figure \ref{fig:angularities}. The plot also shows the analytic thrust results, which we reproduce for $A=0$ accurately. We have further performed four-dimensional fits to the data points that allow to extract the two-loop coefficients for inter\-mediate values of the angularity. Our calculation provides the last missing ingredient for NNLL resummation of the angularity distributions. 

\section{Outlook}

\begin{figure}[t]
\centering
\hspace{2mm}
\includegraphics[scale=0.55]{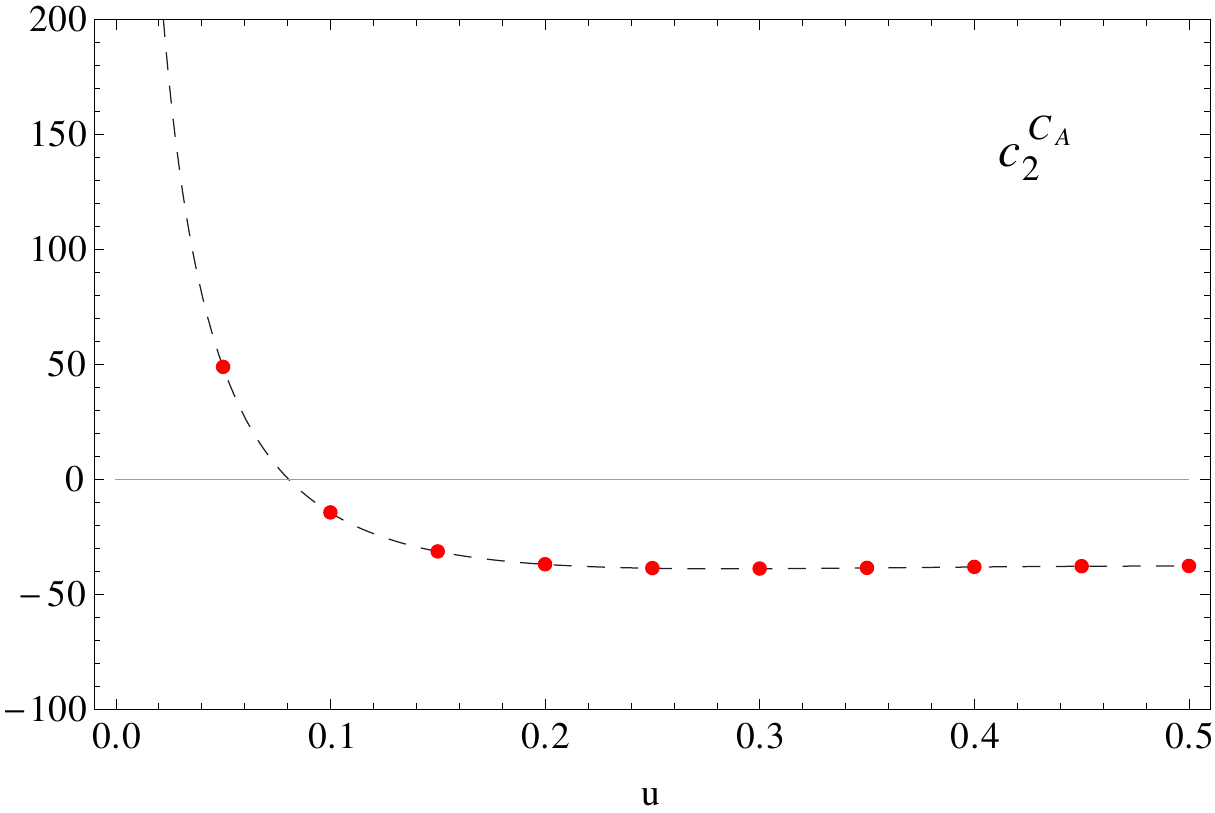}
\hspace{8mm}
\includegraphics[scale=0.55]{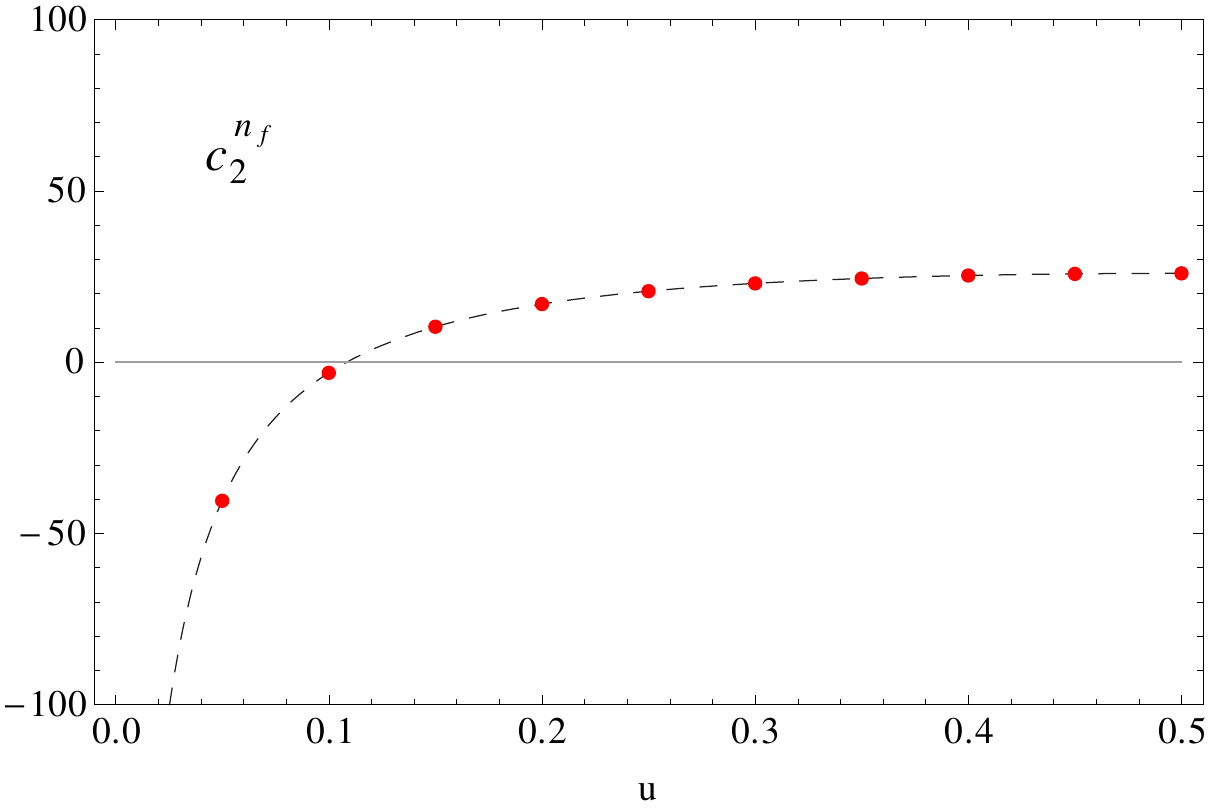}
\caption{Two-loop finite term of the renormalised hemisphere soft function. The dashed line represents the analytic result of \cite{Kelley:2011ng}.}
\label{fig:hemispheremasses}
\end{figure}

Our algorithm applies to dijet soft functions that are defined with a measurement function of the form (\ref{eq:measure:NLO}) at NLO and (\ref{eq:measure:NNLO}) at NNLO. We have shown that this encapsulates a large class of soft functions, which have typically been calculated analytically on a case-by-case basis in the past. In the future we plan to extend the range of observables in several respects:
\begin{itemize}
  \item
  We assumed that the measurement function depends linearly on $k_T$ since the Laplace variable $\tau$ has the dimension $1/$mass. This can easily be generalised to arbitrary powers, which would only modify the analytic part of the calculation.
  \item
  So far we only considered measurement functions that are consistent with non-abelian exponentiation, see (\ref{eq:measure:NAE}). It should be possible to calculate the $C_F^2$ contribution with similar techniques, which would make this limitation obsolete.
  \item
For simplicity we focused on soft functions that depend on a single Laplace variable $\tau$. Multi-differential soft functions can also be accommodated in our approach, and as an example we show the results for the two-loop finite term of the hemisphere soft function in Figure \ref{fig:hemispheremasses}. 
The result is shown as a function of the variable $u=\tau_L/(\tau_L+\tau_R)$, where $\tau_L$ and $\tau_R$ are the respective Laplace variables.   
  \item
  We saw in (\ref{eq:NLO:master}) that the rapidity integral diverges for $n=0$. This  corresponds to SCET-2 type observables, which need an additional regulator on top of DR. Our algorithm can easily be adopted to include the rapidity regulator of \cite{Becher:2011dz}, but the current version of {\tt SecDec} is designed for only one type of infrared regulator. The implementation of a second regulator is in development, which will immediately allow calculations for SCET-2 observables.
  \item
  It is sometimes preferable to consider cumulant instead of Laplace (or Fourier) space soft functions. In this case the exponential in the measurement function is replaced by a step function, which again would only modify the analytic structure of our calculation. Alternatively, one could bring the cumulant measurement function into an exponentiated form by applying an additional Laplace transformation.  
  \item
  There are other observables which do not fall into the class considered here. An example is the jet broadening soft function, which depends on the total transverse momentum of the soft radiation due to recoil effects. The broadening soft function is therefore conveniently discussed in a combined Laplace-Fourier space, and the measurement function contains an additional $\eps$-dependent function from the $(d-2)$-dimensional Fourier transformation \cite{Becher:2011pf}. As this function is independent of the singularity structure, it can be expanded in $\eps$ before performing the phase-space integrations, and it should therefore not lead to additional complications. 
\end{itemize}


\begin{thebibliography}{99} 

\bibitem{Bauer:2000yr}
  C.~W.~Bauer, S.~Fleming, D.~Pirjol and I.~W.~Stewart,
  Phys.\ Rev.\ D {\bf 63} (2001) 114020
  [hep-ph/0011336].
  
\bibitem{Bauer:2001yt}
  C.~W.~Bauer, D.~Pirjol and I.~W.~Stewart,
  Phys.\ Rev.\ D {\bf 65} (2002) 054022
  [hep-ph/0109045].
  
\bibitem{Beneke:2002ph}
  M.~Beneke, A.~P.~Chapovsky, M.~Diehl and T.~Feldmann,
  Nucl.\ Phys.\ B {\bf 643} (2002) 431
  [hep-ph/0206152].
  
\bibitem{Belitsky:1998tc}
  A.~V.~Belitsky,
  Phys.\ Lett.\ B {\bf 442} (1998) 307
  [hep-ph/9808389].
  
\bibitem{Kelley:2011ng}
  R.~Kelley, M.~D.~Schwartz, R.~M.~Schabinger and H.~X.~Zhu,
  Phys.\ Rev.\ D {\bf 84} (2011) 045022
  [arXiv:1105.3676 [hep-ph]].
  
\bibitem{Monni:2011gb}
  P.~F.~Monni, T.~Gehrmann and G.~Luisoni,
  JHEP {\bf 1108} (2011) 010
  [arXiv:1105.4560 [hep-ph]].
  
\bibitem{Li:2011zp}
  Y.~Li, S.~Mantry and F.~Petriello,
  Phys.\ Rev.\ D {\bf 84} (2011) 094014
  [arXiv:1105.5171 [hep-ph]].
  
\bibitem{Kelley:2011aa}
  R.~Kelley, M.~D.~Schwartz, R.~M.~Schabinger and H.~X.~Zhu,
  Phys.\ Rev.\ D {\bf 86} (2012) 054017
  [arXiv:1112.3343 [hep-ph]].
  
\bibitem{Becher:2012za}
  T.~Becher, G.~Bell and S.~Marti,
  JHEP {\bf 1204} (2012) 034
  [arXiv:1201.5572 [hep-ph]].
  
\bibitem{Ferroglia:2012uy}
  A.~Ferroglia, B.~D.~Pecjak, L.~L.~Yang, B.~D.~Pecjak and L.~L.~Yang,
  JHEP {\bf 1210} (2012) 180
  [arXiv:1207.4798 [hep-ph]].
  
\bibitem{Becher:2012qc}
  T.~Becher and G.~Bell,
  JHEP {\bf 1211} (2012) 126
  [arXiv:1210.0580 [hep-ph]].
  
\bibitem{Czakon:2013hxa}
  M.~Czakon and P.~Fiedler,
  Nucl.\ Phys.\ B {\bf 879} (2014) 236
  [arXiv:1311.2541 [hep-ph]].
  
\bibitem{vonManteuffel:2014mva}
  A.~von Manteuffel, R.~M.~Schabinger and H.~X.~Zhu,
  Phys.\ Rev.\ D {\bf 92} (2015) 4,  045034
  [arXiv:1408.5134 [hep-ph]].
  
\bibitem{Boughezal:2015eha}
  R.~Boughezal, X.~Liu and F.~Petriello,
  Phys.\ Rev.\ D {\bf 91} (2015) 9,  094035
  [arXiv:1504.02540 [hep-ph]].
  
\bibitem{Echevarria:2015byo}
  M.~G.~Echevarria, I.~Scimemi and A.~Vladimirov,
  arXiv:1511.05590 [hep-ph].
  
\bibitem{Becher:2014aya}
  T.~Becher, R.~Frederix, M.~Neubert and L.~Rothen,
  Eur.\ Phys.\ J.\ C {\bf 75} (2015) 4,  154
  [arXiv:1412.8408 [hep-ph]].
  
\bibitem{Banfi:2004yd}
  A.~Banfi, G.~P.~Salam and G.~Zanderighi,
  JHEP {\bf 0503} (2005) 073
  [hep-ph/0407286].
  
\bibitem{Banfi:2014sua}
  A.~Banfi, H.~McAslan, P.~F.~Monni and G.~Zanderighi,
  JHEP {\bf 1505} (2015) 102
  [arXiv:1412.2126 [hep-ph]].
  
\bibitem{Binoth:2000ps}
  T.~Binoth and G.~Heinrich,
  Nucl.\ Phys.\ B {\bf 585} (2000) 741
  [hep-ph/0004013].

\bibitem{Becher:2011dz}
  T.~Becher and G.~Bell,
  Phys.\ Lett.\ B {\bf 713} (2012) 41
  [arXiv:1112.3907 [hep-ph]].

\bibitem{Fleming:2007qr}
  S.~Fleming, A.~H.~Hoang, S.~Mantry and I.~W.~Stewart,
  Phys.\ Rev.\ D {\bf 77} (2008) 074010
  [hep-ph/0703207].

\bibitem{Hornig:2009vb}
  A.~Hornig, C.~Lee and G.~Ovanesyan,
  JHEP {\bf 0905} (2009) 122
  [arXiv:0901.3780 [hep-ph]].

\bibitem{Larkoski:2014uqa}
  A.~J.~Larkoski, D.~Neill and J.~Thaler,
  JHEP {\bf 1404} (2014) 017
  [arXiv:1401.2158 [hep-ph]].

\bibitem{Hoang:2014wka}
  A.~H.~Hoang, D.~W.~Kolodrubetz, V.~Mateu and I.~W.~Stewart,
  Phys.\ Rev.\ D {\bf 91} (2015) 9,  094017
  [arXiv:1411.6633 [hep-ph]].

\bibitem{Korchemsky:1993uz}
  G.~P.~Korchemsky and G.~Marchesini,
  Phys.\ Lett.\ B {\bf 313} (1993) 433.
  
\bibitem{Becher:2009th}
  T.~Becher and M.~D.~Schwartz,
  JHEP {\bf 1002} (2010) 040
  [arXiv:0911.0681 [hep-ph]].
  
\bibitem{Becher:2015gsa}
  T.~Becher and X.~Garcia i Tormo,
  JHEP {\bf 1506} (2015) 071
  [arXiv:1502.04136 [hep-ph]].
  
\bibitem{Gatheral:1983cz}
  J.~G.~M.~Gatheral,
  Phys.\ Lett.\ B {\bf 133} (1983) 90.
    
\bibitem{Frenkel:1984pz}
  J.~Frenkel and J.~C.~Taylor,
  Nucl.\ Phys.\ B {\bf 246} (1984) 231.
  
\bibitem{Carter:2010hi}
  J.~Carter and G.~Heinrich,
  Comput.\ Phys.\ Commun.\  {\bf 182} (2011) 1566
  [arXiv:1011.5493 [hep-ph]].
  
\bibitem{Borowka:2012yc}
  S.~Borowka, J.~Carter and G.~Heinrich,
  Comput.\ Phys.\ Commun.\  {\bf 184} (2013) 396
  [arXiv:1204.4152 [hep-ph]].
  
\bibitem{Borowka:2015mxa}
  S.~Borowka, G.~Heinrich, S.~P.~Jones, M.~Kerner, J.~Schlenk and T.~Zirke,
  Comput.\ Phys.\ Commun.\  {\bf 196} (2015) 470
  [arXiv:1502.06595 [hep-ph]].
  
\bibitem{Hahn:2004fe}
  T.~Hahn,
  Comput.\ Phys.\ Commun.\  {\bf 168} (2005) 78
  [hep-ph/0404043].
    
\bibitem{Kawabata:1995th}
  S.~Kawabata,
  Comput.\ Phys.\ Commun.\  {\bf 88} (1995) 309.
  
  \bibitem{BRT}
  G.~Bell, R.~Rahn and J.~Talbert, in preparation.
  
\bibitem{Becher:2011pf}
  T.~Becher, G.~Bell and M.~Neubert,
  Phys.\ Lett.\ B {\bf 704} (2011) 276
  [arXiv:1104.4108 [hep-ph]].
  
	
\end{thebibliography}
\end{document}